\documentclass[aps,prb, amsmath,superscriptaddress,longbibliography,notitlepage]{revtex4-2}
\usepackage[latin9]{inputenc}
\usepackage{verbatim}
\usepackage{amsmath}
\usepackage{amssymb}
\usepackage{graphicx}
\usepackage{xcolor}
\usepackage{babel}

\begin{document}
\author{Zhuo Bin Siu}
\email{elesiuz@nus.edu.sg}
\affiliation{Department of Electrical and Computer Engineering, National University of Singapore, Singapore 117583, Republic of Singapore}
\author{Anirban Kundu}
\email{anirbank5@gmail.com}
\affiliation{Department of Electrical and Computer Engineering, National University of Singapore, Singapore 117583, Republic of Singapore}
\affiliation{Department of Physics, Ariel University, Ariel 40700, Israel}
\author{Mansoor B.A. Jalil}
\email{elembaj@nus.edu.sg}
\affiliation{Department of Electrical and Computer Engineering, National University of Singapore, Singapore 117583, Republic of Singapore}

\title{Second-order charge and spin transport in LaO/STO system in the presence of cubic Rashba spin orbit couplings}

\begin{abstract}
Certain non-centrosymmetric materials with broken time-reversal symmetry may exhibit non-reciprocal transport behavior under an applied electric field in which the charge and spin currents contain components that are second order in the electric field. In this study, we investigate the second-order spin accumulation and charge and spin responses in the LaAlO$_3$/SrTiO$_3$ (LaO/STO) system with magnetic dopants under the influence of linear and cubic Rashba spin--orbit coupling (RSOC) terms. We explain the physical origin of the second-order response and perform a symmetry analysis of the first and second-order responses for different dopant magnetization directions relative to the applied electric field. We then numerically solve the Boltzmann transport equation by extending the approach of Schliemann and Loss [Phys. Rev. B 68, 165311] to higher orders in the electric field. We show that the sign of the second-order responses can be switched by varying the magnetization direction of the magnetic dopants or relative strengths of the two cubic RSOC terms and explain these trends by considering the Fermi surfaces of the respective systems. These findings provide insights into the interplay of multiple SOC effects in a LaO/STO system and how the resulting first- and second-order charge and spin responses can be engineered by exploiting the symmetries of the system.
\end{abstract}

\maketitle

\section{Introduction}

A characteristic of non-centrosymmetric materials with broken time-reversal symmetry is non-reciprocal transport behavior in which charge currents flowing along opposite directions experience different electrical resistivities. Such non-reciprocal transport has been experimentally observed in, e.g., the bulk polar semiconductor BiTeBr \cite{NatPhy13_578} and the transition metal dichalcogenide WTe$_2$ \cite{NatComm10_1290} and can be interpreted as an electrical resistance that is itself linearly dependent on the applied electric field, thus giving rise to a quadratic dependence of the current on the electric field. This second-order response can be exploited to characterize the surface spin texture \cite{NatPhys14_495,PRL120_266802} or the strength of the Rashba spin-orbit interaction in a material \cite{NatPhy13_578}. Such second-order responses can be generalized to other transport quantities aside from the longitudinal current. For instance, a second-order Hall current perpendicular to the electric field has been observed in WTe$_2$ \cite{NatMat18_324}. Additionally, a second-order response in the spin current has also been demonstrated in the presence of an applied AC field, which when coupled with the absence of a corresponding first-order response in the charge current results in the generation of pure spin current without an accompanying charge current \cite{PRL113_156603, PRL123_016801},

In particular, large second-order responses for the charge current have been experimentally observed in the LaAlO$_3$/SrTiO$_3$ (LaO/STO)  system \cite{NatComm10_4510, PhyRevMat4_071001,PRB114_045114}. The LaO/STO system is a promising candidate material for spintronics applications because of the strong spin-orbit coupling at its interface, which gives rise to a large (first-order) spin-to-charge conversion efficiency  \cite{Wang2017,Trier2020,EPL116_17006}. Besides the usual $\alpha(\boldsymbol{k}\times\boldsymbol{\sigma})\cdot\hat{z}$ linear Rashba spin--orbit coupling (SOC) term, there is also a significant cubic-momentum Rashba SOC at the LaO/STO interface \cite{PhysRevB.93.045108,PhysRevB.92.075309,PhysRevLett.108.206601}. In the LaO/STO system, the linear RSOC has a strength of approximately 10 meV\AA and that of the cubic RSOC is approximately 1--5 eV\AA$^3$ \cite{PhysRevLett.104.126803,PhysRevB.86.201105,PhysRevB.92.075309,Ho_2019}. When considered over the Brillouin zone, the linear and cubic SOC energy splits in the LaO/STO system are comparable in magnitude. 

Recently,  Ho et al. showed that the low-energy Hamiltonian for the LaO/STO system can be described as comprising the usual linear Rashba SOC and two distinct types of cubic Rashba SOC terms. The magnitudes of the three SOC terms can be externally tuned by modulating the thickness of the two-dimensional electron gas at the LaO/STO interface and the out-of-plane electric field \cite{Ho_2019}. In general, the cubic SOC terms play a key role in inducing non-linear spin and current responses. For example, it has been predicted that the trigonal warping term in transition metal dichalogenides can give rise to non-linear spin current \cite{PRL113_156603}, whereas in Bi$_2$Te$_3$, which is a topological insulator,  the experimentally observed nonlinear magnetoresistance \cite{NatPhys14_495} and Hall current \cite{PRL123_016801} were attributed to the hexagonal warping  \cite{JPD49_225304,AIPAdv6_055706, SciRep4_5062} of its Fermi surface due to the cubic SOC terms.  However, to the best of our knowledge, the effects of these cubic Rashba terms on the second-order transport phenomena such as the transverse charge current and spin accumulation have not yet been studied. 

Therefore, in this study, we investigate the second-order charge current, spin accumulation, and spin current in the LaO/STO system in the presence of the linear and the two distinct cubic Rashba SOC terms. We apply the Boltzmann transport model and solve it to the second order in the applied electric field. We adopt the more refined scattering model of  Schliemann and Loss \cite{PhysRevB.68.165311} (SL), which yields a more accurate representation up to the higher orders compared to the relaxation time approximation (RTA). The SL approach is first extended to higher orders in the applied electric field and then used to solve the Boltzmann model. We elucidate the physical origin of the second-order charge and spin transport responses by considering the symmetry of the first- and second-order responses. Based on the symmetry analysis, we explain how the second-order responses vary as functions of the strengths of the various SOC terms as well as the magnetization coupling.  

\section{Extension of Schliemann-Loss approach}

	The Boltzmann equation for a time-invariant homogeneous system with an applied electric field $\boldsymbol{E}$ is given by
	\begin{equation}
		e\boldsymbol{E}\cdot\partial_{\boldsymbol{k}} f_\mu(\boldsymbol{k}) = \partial_t f_\mu(\boldsymbol{k}), \label{be1}
	\end{equation}
	where $f_\mu(\boldsymbol{k})$ is the occupation function of the $\mu$th band.
	
	We assume that $\partial_t f_\mu(\boldsymbol{k})$ has the general form of 
	\begin{align}
		\partial_t f_\mu(\boldsymbol{k}) &= -S[f_\mu(\boldsymbol{k})]  \nonumber \\
	\end{align}
where $S[f_\mu]$ denotes that $S$ is a functional of $f_\mu$. In the commonly adopted RTA approach, $S[f_\mu]$ is assumed to have the form of $S[f_\mu] = (f_\mu-f^{(0)})/\tau$ where $\tau$ is a constant relaxation time and $f^{(0)}$ is the Fermi--Dirac distribution. Subsequently, a more refined approximation for the scattering functional was adopted by Schleimann and Loss who assumed a collision
integral in the form of \cite{0030839939} 
\begin{equation}
S[f_{\boldsymbol{k}}] =\int \mathrm{d}\boldsymbol{k'} \sum_{\mu,\mu'}w_{\boldsymbol{k}\mu,\boldsymbol{k'}\mu'}\left(f_\mu(\boldsymbol{k})-f_{\mu'}(\boldsymbol{k'})\right) \label{eq:SLIcoll1}
\end{equation}
where $w_{\boldsymbol{k},\mu,\boldsymbol{k}',\mu'}$ is the transition probability for electrons in the $(\boldsymbol{k},\mu)$ state to be scattered into the $(\boldsymbol{k}', \mu')$ state.  
We consider electrons to be scattered by impurities,  each of which is modeled as a delta
function potential 
\begin{equation}
	V_i(\boldsymbol{r})=a_0V_{0}\delta(\boldsymbol{r}-\boldsymbol{R}_{i}) \label{eq:Vr}
\end{equation}
where $\boldsymbol{R}_{i}$ is the spatial location of the impurity, $V_{0}$ its scattering strength, and $a_0$ is a quantity with the physical dimensions of area introduced for dimensional consistency \cite{Paper1}. Denoting the number density (i.e., the number per unit area) of such impurities as $n$ and applying the Fermi golden rule, the transition rate due to all the impurities is given by  
\begin{equation}
w_{\boldsymbol{k},\mu,\boldsymbol{k}',\mu'}  = 2\pi n\left|\langle\boldsymbol{k},\mu|a_0V_0|\boldsymbol{k}',\mu'\rangle\right|^{2}\delta\left(\epsilon_\mu(\boldsymbol{k})-\epsilon_{\mu'}(\boldsymbol{k'})\right). \label{eq:wkkprime1}
\end{equation}

In analogy to the Schliemann and Loss approach \cite{PhysRevB.68.165311}, let $\theta_\mu(\boldsymbol{k})$ be the angle between the direction of the velocity $\langle \boldsymbol{v} \rangle$ for the state at $\boldsymbol{k}$ on the $\mu$th band and the applied electric field $\boldsymbol{E}$. Consider 
\begin{align}
	& S[v_\mu(\boldsymbol{k}) \exp(i\theta_\mu(\boldsymbol{k}))] \nonumber \\
	&=  \sum_{\mu'} \int \frac{\mathrm{d}S_{k'}}{2\pi^2}  (v_\mu(\boldsymbol{k})\exp(i\theta_\mu(\boldsymbol{k})) - v_{\mu'}(\boldsymbol{k'})\exp(i\theta_{\mu'}(\boldsymbol{k'}))) w_{\boldsymbol{k},\mu,\boldsymbol{k}',\mu'} \\
	&=  \left[ \sum_{\mu'} \int \frac{\mathrm{d}S_{k'}}{2\pi^2}  \left(1 - \frac{ v_{\mu'}(\boldsymbol{k'})}{v_\mu(\boldsymbol{k})}\exp(i(\theta_{\mu'}(\boldsymbol{k'})-\theta_{\mu}(\boldsymbol{k})))\right)w_{\boldsymbol{k},\mu,\boldsymbol{k}',\mu'} \right] v_\mu(\boldsymbol{k})\exp(i\theta_\mu(\boldsymbol{k})).
\end{align}

By comparing the real and imaginary parts, we have
\begin{align}
	&S[v_\mu(\boldsymbol{k})\cos(\theta_\mu(\boldsymbol{k}))] \nonumber \\
	=& \sum_{\mu'}   \int \frac{\mathrm{d}S_k}{(2\pi)^2}  \left[ - \left( \frac{ v_{\mu'}(\boldsymbol{k'})}{v_\mu(\boldsymbol{k})} \sin(\theta_{\mu'}(\boldsymbol{k'})-\theta_{\mu}(\boldsymbol{k})) \right)  v_\mu(\boldsymbol{k})\sin(\theta_\mu(\boldsymbol{k})) \right. \nonumber \\
	&+ \left.  \left(1 - \frac{ v_{\mu'}(\boldsymbol{k'})}{v_\mu(\boldsymbol{k})}\cos(\theta_{\mu'}(\boldsymbol{k'}))\right) v_\mu(\boldsymbol{k})\cos(\theta_\mu(\boldsymbol{k}))\right]w_{\boldsymbol{k},\mu,\boldsymbol{k}',\mu'}, \\
	&S[v_\mu(\mathbf{k})\sin(\theta_\mu(\mathbf{k})] \nonumber \\
	=& \sum_{\mu'}   \int \frac{\mathrm{d}S_k}{(2\pi)^2} \left[ - \left( \frac{ v_{\mu'}(\mathbf{k'})}{v_\mu(\mathbf{k})} \sin(\theta_{\mu'}(\mathbf{k'})-\theta_{\mu}(\mathbf{k})) \right) v_\mu(\mathbf{k})\cos(\theta_\mu(\mathbf{k})) \right. \nonumber \\
	&+ \left.  \left(1 - \frac{ v_{\mu'}(\mathbf{k'})}{v_\mu(\mathbf{k})}\cos(\theta_{\mu'}(\mathbf{k'}))\right) v_\mu(\mathbf{k})\sin(\theta_\mu(\mathbf{k}))\right]w_{\boldsymbol{k},\mu,\boldsymbol{k}',\mu'}. 
\end{align}
We define 
\begin{align}
	(\tau^\parallel_\mu(\boldsymbol{k}))^{-1} &= \sum_{\mu'} \int \frac{\mathrm{d}S_{\mathrm{k}}}{(2\pi)^2} w_{\boldsymbol{k},\mu,\boldsymbol{k}',\mu'}\left(1 - \frac{ v_{\mu'}(\boldsymbol{k'})}{v_\mu(\boldsymbol{k})}\cos(\theta_{\mu'}(\boldsymbol{k'})-\theta_{\mu}(\boldsymbol{k}))\right)\\
	(\tau^\perp_\mu(\boldsymbol{k}))^{-1} &= \sum_{\mu'} \int \frac{\mathrm{d}S_{\mathrm{k}}}{(2\pi)^2} w_{\boldsymbol{k},\mu,\boldsymbol{k}',\mu'}\left( \frac{ v_{\mu'}(\boldsymbol{k'})}{v_\mu(\boldsymbol{k})} \sin(\theta_{\mu}(\boldsymbol{k})-\theta_{\mu'}(\boldsymbol{k}')) \right), 
\end{align}
so that the scattering functionals can be written as
\begin{equation}
	S \begin{bmatrix}  v_\mu(\boldsymbol{k}) \cos(\theta_\mu(\boldsymbol{k})) \\  v_\mu(\boldsymbol{k}) \sin(\theta_\mu(\boldsymbol{k})) \end{bmatrix} = \begin{pmatrix} \tau^\parallel_\mu(\boldsymbol{k})^{-1} & -\tau^\perp_\mu(\boldsymbol{k})^{-1} \\ \tau^\perp_\mu(\boldsymbol{k})^{-1} & \tau^\parallel_\mu(\boldsymbol{k})^{-1} \end{pmatrix} \begin{pmatrix} v_\mu(\boldsymbol{k}) \cos(\theta_\mu(\boldsymbol{k})) \\ v_\mu(\boldsymbol{k}) \sin(\theta_\mu(\boldsymbol{k})) \end{pmatrix}. \label{Smat}
\end{equation}

To extend the SL formalism to higher orders in $\boldsymbol{E}$, we expand $f_\mu(\boldsymbol{k})$ as a series in $E=|\boldsymbol{E}|$, i.e.,  $f_\mu(\boldsymbol{k}) = \sum_n f_\mu ^{(n)}(\boldsymbol{k})E^n$. 

To the first order in $E$, Eq. \eqref{be1} gives
\begin{equation}
	(e\boldsymbol{E}\cdot\partial_{\boldsymbol{k}}\epsilon) \partial_\epsilon f^{(0)}(\boldsymbol{k}) = -S[f^{(1)}_\mu(\boldsymbol{k}) E] \label{f1}.
\end{equation} 

We apply the ansatz 
\begin{equation}
	f^{(1)}_\mu(\boldsymbol{k}) E = eE(\partial_\epsilon f^{(0)}(\boldsymbol{k}))(c^{(1)} v_\mu(\boldsymbol{k})\cos(\theta_\mu(\boldsymbol{k})) + s^{(1)} v_\mu(\boldsymbol{k})\sin(\theta_\mu(\boldsymbol{k})) ) \label{f1ansatz}
\end{equation} 
where $c^{(1)}$ and $s^{(1)}$ are unknown coefficients to be determined. Based on Eq. \eqref{Smat}, Eq. \eqref{f1} reduces to (we drop the $\boldsymbol{k}$ arguments here for notational simplicity)
\begin{equation}
	-\cos(\theta_\mu) = c^{(1)} \left((\tau^\parallel_\mu)^{-1}\cos(\theta_\mu)-(\tau^\perp_\mu)^{-1}\sin(\theta_\mu)\right) + s^{(1)} \left((\tau^\perp_\mu)^{-1}\cos(\theta_\mu) + (\tau^\parallel_\mu)^{-1}\sin(\theta_\mu)\right). 
\end{equation}

Comparing the coefficients of $\cos(\theta_\mu)$ and $\sin(\theta_\mu)$ on both sides of the equation, we have 
\begin{align}
	\left( c^{(1)}(\tau^\parallel_\mu)^{-1}+ s^{(1)}(\tau^\perp_\mu)^{-1}\right) &= -1, \\
	\left( c^{(1)}(-\tau^\perp_\mu)^{-1}+ s^{(1)}(\tau^\parallel_\mu)^{-1}\right) &= 0. \label{c1s1eq} 
\end{align}

This gives
\begin{align}
	c^{(1)} &= -\frac{\tau^\parallel_\mu}{1 + \left( \frac{ \tau^\parallel_\mu}{\tau^\perp_\mu} \right)^2 } \\
	s^{(1)} &= -\frac{\tau^\perp_\mu }{1 + \left( \frac{ \tau^\perp_\mu}{\tau^\parallel_\mu} \right)^2 }.
\end{align}

The second-order terms in $\boldsymbol{E}$ in Eq. \eqref{be1} read
\begin{align}
	& e\boldsymbol{E}\cdot\partial_{\boldsymbol{k}}f^{(1)}_\mu E_\mu = -S[f^{(2)}_\mu E^2], \\
	& \Rightarrow ev_\mu\cos(\theta_\mu) (\partial_\epsilon f^{(1)}_\mu ) = -S[f^{(2)}_\mu ]. \label{f2}
\end{align}

In analogy to Eqs. \eqref{f1} and \eqref{f1ansatz}, this suggests that we can adopt a similar ansatz in Eq. \eqref{f2}:
\begin{equation}
	f^{(2)}_\mu = e(\partial_\epsilon f^{(1)}) (c^{(2)} v_\mu(\boldsymbol{k})\cos(\theta_\mu(\boldsymbol{k})) + s^{(2)} v_\mu(\boldsymbol{k})\cos(\theta_\mu(\boldsymbol{k}))
\end{equation}
This gives an analogous set of equations to Eq. \eqref{c1s1eq} with the $c^{(1)}$ and $s^{(1)}$ replaced by $c^{(2)}$ and $s^{(2)}$, respectively. $c^{(2)}$ and $s^{(2)}$ therefore have the same forms as $c^{(1)}$ and $s^{(1)}$, respectively. 

Generalizing the above results, we see that the $n$th order term in the expansion of the occupation function is given by 
\begin{equation}
	Ef^{(n)}_\mu(\boldsymbol{k}) = -\left[e Ev_\mu(\boldsymbol{k})\left( \frac{\tau^\parallel_\mu(\boldsymbol{k})}{1 + \left( \frac{ \tau^\parallel_\mu(\boldsymbol{k})}{\tau^\perp_\mu(\boldsymbol{k})} \right)^2 }\cos(\theta_\mu(\boldsymbol{k})) + \frac{\tau^\perp_\mu(\boldsymbol{k})}{1 + \left( \frac{ \tau^\perp_\mu(\boldsymbol{k})}{\tau^\parallel_\mu(\boldsymbol{k})} \right)^2 }\sin(\theta_\mu(\boldsymbol{k}))\right)\right] (\partial_\epsilon f^{(n-1)}_\mu(\boldsymbol{k})). \label{SLfn}
\end{equation}
In contrast to the conventional RTA in which a constant $\boldsymbol{k}$-independent relaxation time $\tau$ is used for all states on all bands, the scattering in the SL approach now differs for each state. Moreover, the SL approach allows for the possibility that change in momentum may be perpendicular to the direction of the electric field as denoted by the $\sin(\theta_\mu(\boldsymbol{k}))$ terms, unlike the RTA in which the change in momentum is restricted to being parallel to the applied electric field. 

\subsection{Physical interpretation of first- and second-order terms} 

Because the terms in the large round brackets in Eq. \eqref{SLfn} have the dimensions of time, their product with $e E$ have the dimensions of $\boldsymbol{k}$. This product can thus be interpreted as the $\boldsymbol{k}$-space shift of the Fermi surface induced by the electric field. The terms in the square brackets in Eq. \eqref{SLfn} in turn have the dimensions of energy and can be interpreted as the energy shift induced by the shift in the Fermi surface. Denoting these terms as $\delta \epsilon_\mu$ and introducing $\delta f^{(n)}_\mu \equiv E^n f^{(n)}_\mu$, we have, explicitly, 
\begin{align}
	\delta f^{(1)}_\mu = -(\delta\epsilon_\mu)(\partial_\epsilon f^{(0)}_\mu), \label{Ef1}  \\
	\delta f^{(2)}_\mu = -(\delta\epsilon_\mu)(\partial_\epsilon \delta f^{(1)}_\mu). \label{Ef2} 
\end{align} 

We note that Eqs. \eqref{Ef1} and \eqref{Ef2} are also applicable to the conventional RTA approach, for which $\delta\epsilon_\mu=v_\mu\cos(\theta_\mu)E\tau$ where $\tau$ is the RTA relaxation time. In this section and the next sections, we base our explanation on the RTA for expositional simplicity although the arguments can be readily extended to the SL approach by using the appropriate $\delta\epsilon_\mu$ given in Eq. \eqref{SLfn}. Figure S1 in the Supplemental Materials show that the RTA and SL approaches produce qualitatively similar results except for a small underestimation of the magnitudes of the second-order responses by the RTA approach.

\begin{figure}[htp]
\centering
\includegraphics[width=0.42\textwidth]{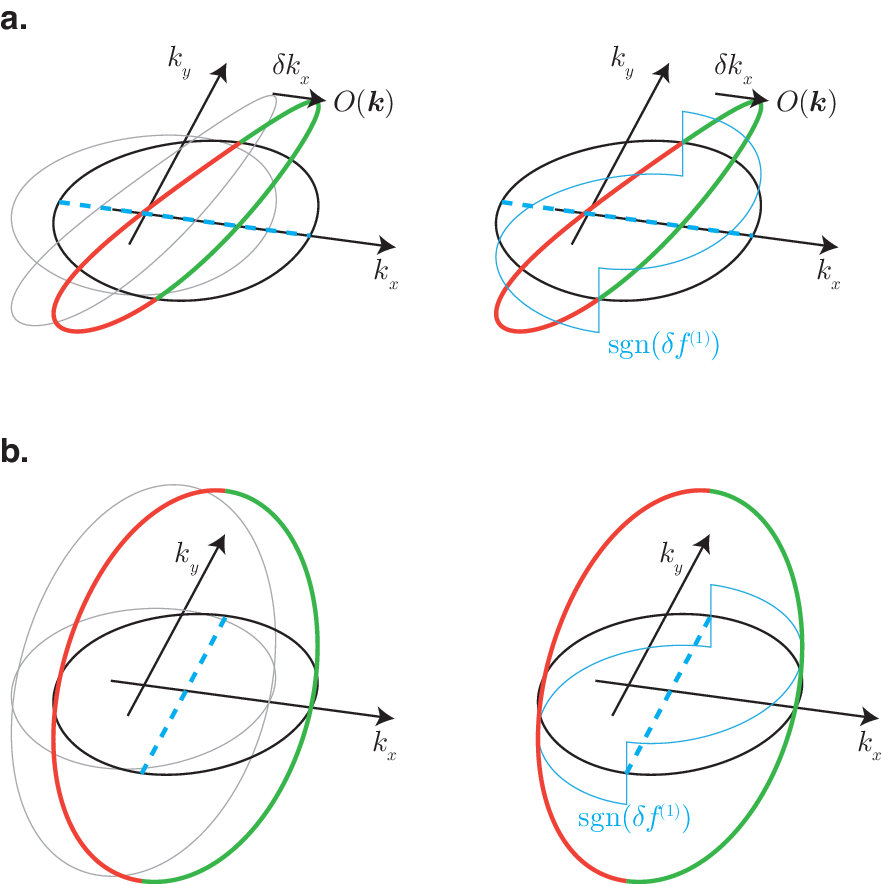}
\caption{A schematic illustration of the first-order change $\delta O^{(1)}$ for (a) the observable $O=v_x$ in a system coupled  to a magnetization in the $y$ direction, for which the Fermi surface is symmetrical about the $k_x$ axis,  and (b) the observable $O=v_y$ in a system coupled to a magnetization in the $x$ direction, for which the Fermi surface is symmetrical about the $k_y$ axis. The symmetry axes of the systems are denoted by the blue dotted lines along the $k_x$ or $k_y$ axes. The lighter gray ellipsoids on the $k_x-k_y$ plane in the left plots represent the original Fermi surfaces, the black ellipsoids are the Fermi surfaces shifted by $\delta k_x$ due to an applied electric field in the $x$ direction, and the lighter gray curves the values of $O(\boldsymbol{k})$ on the original Fermi surface. The blue half-ellipsoids denoted as $\mathrm{sgn} (\delta f^{(1)})$ in the right plots denote the sign of the first-order change in the Fermi-Dirac occupancy factor where the portions above (below) the $k_x-k_y$ plane denote a positive (negative) value.  The portions of $O(\boldsymbol{k})$ in green (red) are multiplied by positive (negative) values of $\delta f^{(1)}$ in the expressions for $\delta O^{(1)}$. }
\label{gFsym1o}
\end{figure}

The first-order response $\delta O^{(1)}$  of an observable quantity $O$ (e.g., the $x$ and $y$ velocities $v_{x,y}$ or spin polarizations $\sigma_{x,y}$ )  to the electric field is given by 
\begin{equation}
	\delta O^{(1)} = \int \mathrm{d}\boldsymbol{k}\ \left( \sum_\mu \delta f^{(1)}_\mu(\boldsymbol{k}) O_\mu(\boldsymbol{k}) \right) \label{O1} 
\end{equation}
where  $O_\mu(\boldsymbol{k})\equiv \langle \boldsymbol{k};\mu|O|\boldsymbol{k};\mu\rangle$ is the expectation value of the quantity. 
Substituting Eq. \eqref{Ef1} into Eq. \eqref{O1}, the $\delta f^{(1)}_\mu(\boldsymbol{k}) O_\mu(\boldsymbol{k})$ term  becomes $(-\partial_\epsilon f^{(0)}_\mu\delta\epsilon_\mu)O_\mu(\boldsymbol{k})$. Noting that $\partial_\epsilon f^{(0)}_\mu(\boldsymbol{k})=-\delta(\epsilon_\mu(\boldsymbol{k})-E_f)$ at zero temperature, i.e., the integrand in Eq. \eqref{O1} only has a finite value on the Fermi surface, it becomes evident that $\delta O^{(1)}$ is the change in $O$ due to the first-order change in the occupancy of the states in the vicinity of the original Fermi surfaces caused by the electric-induced shifts of the Fermi surfaces. This is schematically illustrated for $O=v_x$ and an applied electric field along the $x$ direction in Fig. \ref{gFsym1o} for a single band (the band index $\mu$ is omitted in the figure for simplicity). In the RTA, $\delta\epsilon_\mu$ is proportional to $\langle v_x \rangle$. In Fig. \ref{gFsym1o}(a), the portions of $O(\boldsymbol{k})$ multiplied by positive (negative) values of $\delta f^{(1)}$ in Eq. \eqref{O1} corresponding to states with positive (negative) values of $\langle v_x \rangle$ are denoted in green (red). The states on the left half of the Fermi surface with negative velocities parallel to the electric field correspond to previously occupied states which now became unoccupied because their states got scattered into states with more positive values of $k_x$ by the electric field. This loss in occupancy results in a reduction of the total energy of the occupied states on the left half of the Fermi surface and a negative $\delta\epsilon_\mu$ because there are now fewer occupied states. Conversely, the previously unoccupied states with energies slightly higher than the Fermi energy on the right half of the Fermi surface that now become occupied, thereby corresponding to a gain in energy of the occupied states and a positive $\delta\epsilon_\mu$ because there are now more occupied states there.

\begin{figure}[htp]
\centering
\includegraphics[width=0.5\textwidth]{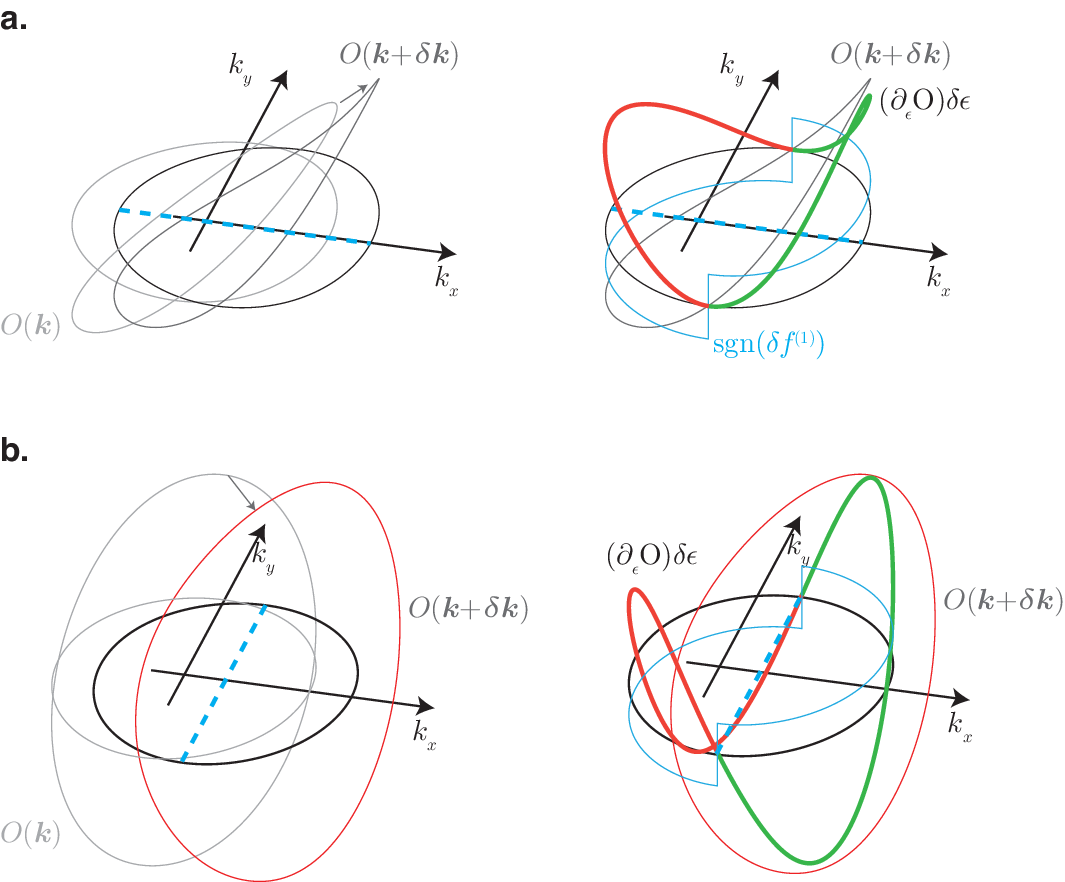}
\caption{A schematic illustration of  the $(\partial_\epsilon O)\delta_\epsilon$ second-order changes in a single band for (a) the observable $O= v_x$ in a system coupled to a magnetization in the $y$ direction, for which the Fermi surface is symmetrical about the $k_x$ axis,  and (b) the observable $O=v_y$ in a system coupled to a magnetization in the $x$ direction, for which the Fermi surface is symmetrical about the $k_y$ axis. The symmetry axes of the systems are denoted by the blue dotted lines along the $k_x$ or $k_y$ axes. The lighter gray ellipsoids in the left plot on the $k_x-k_y$ plane represent the original Fermi surfaces and the lighter gray curves the values of $O(\boldsymbol{k})$ on the original Fermi surface. The black ellipsoids and the darker gray lines in the left and right plots represent the shifted Fermi surfaces and the values of $O(\boldsymbol{k}+\delta\boldsymbol{k})$ on the shifted Fermi surfaces, respectively. The blue half-ellipsoids denoted as $\mathrm{sgn} (\delta f^{(1)})$ on the right plots denote the sign of the first-order change in the Fermi-Dirac occupancy factor where the portions above (below) the $k_x-k_y$ plane denote a positive (negative) value. The portions  of $(\partial_\epsilon O)\delta\epsilon$ in the right plots in green (red) are multiplied by positive (negative) values of $\delta f^{(1)}$ in the expressions for $\delta O^{(1)}$ and $\delta O^{(2)}$ .  }
\label{gFsym2o}
\end{figure}	

The second order change in $O$, $\delta O^{(2)}$, has the following interpretation. Analogous to Eq. \eqref{O1}, $\delta O^{(2)}$ is given by
\begin{equation} 
	\delta O^{(2)} = \sum_\mu \int \mathrm{d}\boldsymbol{k}\ \left(  \delta f^{(2)}_\mu(\boldsymbol{k}) O_\mu(\boldsymbol{k}) \right) \label{O2}.
\end{equation} 
Substituting Eq. \eqref{Ef2} into Eq. \eqref{Ef1} gives 

\begin{align}
	\delta O^{(2)} &= -\sum_\mu \int \mathrm{d}\boldsymbol{k}\ \left(  (\partial_\epsilon\delta f^{(1)}_\mu)(\delta\epsilon_\mu O_\mu) \right) \label{O2a} \\
	&= -\sum_\mu \int \mathrm{d}\boldsymbol{k}\ \left[   \partial_\epsilon(\delta\epsilon_\mu\delta f^{(1)}_\mu O_\mu) - \partial_\epsilon(\delta\epsilon_\mu O_\mu)(\delta f^{(1)}_\mu) \right] \label{O2b}
\\
	&=\sum_\mu \int \mathrm{d}\boldsymbol{k}\  [(\partial_\epsilon\delta\epsilon_\mu) O_\mu + \delta\epsilon_\mu (\partial_\epsilon O_\mu)](\delta f^{(1)}_\mu) \label{O2c}.
\end{align}  
In going from Eq. \eqref{O2b} to Eq. \eqref{O2c}, we made use of the fact that the integration limits of $\mathbf{k}$ span from $\epsilon_\mu=-\infty$ to $\infty$ while the $\partial_\epsilon(\delta\epsilon_\mu\delta f^{(1)}_\mu O_\mu)$ term in the integration contains a factor of $\delta(\epsilon-\epsilon_{\mathrm{F}})$ within  $\delta f^{(1)}_\mu$. This term therefore does not contribute to the integral because the delta function is zero at the integration limits of $\epsilon_\mu$ and can be omitted. The integrand in Eq. \eqref{O2c} contains two terms. The first term $(\partial_\epsilon\delta\epsilon_\mu)\delta f^{(1)}_\mu O_\mu$ accounts for the energy dependence of $\delta\epsilon$ itself (both  $\boldsymbol{v}(\boldsymbol{k})$ in $\delta\epsilon$ and the eigenenergy $\epsilon(\boldsymbol{k})$ are functionally related through their common dependence on $\boldsymbol{k}$). Substituting Eq. \eqref{f1} for $\delta f^{(1)}_\mu$ into the expression gives 
\begin{equation}
	(\partial_\epsilon\delta\epsilon_\mu)\delta f^{(1)}_\mu = -(\partial_\epsilon f^{(0)}_\mu)[(\partial_\epsilon\delta\epsilon_\mu) \delta\epsilon_\mu]. \label{dededf1} 
\end{equation}
The term in the square brackets on the right hand side can be interpreted as a change in energy that is second order in $E$, $\delta\epsilon_\mu^{(2)} \equiv (\partial_\epsilon\delta\epsilon_\mu)\delta\epsilon_\mu$ so that the term in Eq. \eqref{dededf1} can be written as $\delta\epsilon_\mu^{(2)}\partial_\epsilon f^{(0)}_\mu$. This has the physical interpretation of a change in the Fermi-Dirac distribution occupancy factor arising from the second-order shift in energy due to the displacement of the Fermi surface. The second term in the integrand in Eq. \eqref{O2c}, $(\partial_\epsilon O_\mu\delta\epsilon_\mu)(\delta f^{(1)}_\mu)$ is a product of two factors. The first term $(\partial_\epsilon O_\mu\delta\epsilon_\mu)$ corresponds to the change in the expectation value of $O$ at  $\boldsymbol{k}$ on the Fermi surface of the $\mu$th band due to the Fermi surface shift from $\boldsymbol{k}$ to $\boldsymbol{k}+\delta\boldsymbol{k}$ (Fig. \ref{gFsym2o}). The second term $\delta f^{(1)}_\mu$ corresponds to the first-order change in the occupancy factor of the states in the vicinity of the original Fermi surface due to the $k$-space shift of the Fermi surface.  
 
\section{Symmetry analysis} \label{sec:SymAnal}

We apply the general results obtained in the previous section to the specific case of the low-energy Hamiltonian for the lowest-energy pair of spin states in the LaO/STO system. Including the cubic spin-orbit coupling terms, the Hamiltonian  is given by  \cite{Ho_2019}
\begin{equation}
H=\frac{k^{2}}{2m^*}+J_{H}\boldsymbol{\sigma}\cdot\boldsymbol{M}+\alpha(\boldsymbol{\sigma}\times\boldsymbol{k})\cdot\hat{z}+\sigma_{x}\text{\ensuremath{\beta_{3}}}\left(k_{y}k_{x}^{2}-k_{y}^{3}\right)-\sigma_{x}\text{\ensuremath{\eta_{3}}}k_{x}^{2}k_{y}+\sigma_{y}\text{\ensuremath{\beta_{3}}}\left(k_{x}k_{y}^{2}-k_{x}^{3}\right)+\sigma_{y}\text{\ensuremath{\eta_{3}}}k_{y}^{2}k_{x}\label{eq:h-main}
\end{equation}
where $\alpha$ represents the strength of the linear Rashba SOC (RSOC), $\beta_{3}$
and $\eta_{3}$ are coefficients of the two distinct cubic SOC terms, and $J_{H}$
 the Heisenberg exchange interaction between conduction electrons
and the magnetic moments of the dopants $\boldsymbol{M}$. The effects of each of these RSOI terms on the Fermi surfaces and spin polarizations have been described in more detail previously \cite{Paper1}. Here, we consider the effects of coupling to magnetic dopants  that have moments lying in the in-plane direction $\boldsymbol{M}= M(\cos(\phi_{\mathrm{m}})\hat{x} + \sin(\phi_{\mathrm{m}})\hat{y}) = M_x\hat{x} + M_y\hat{y}$.

As before, we assume that the electric field is applied along the $x$ direction. To understand which observables have finite first- and second-order response, we consider the symmetry properties of the LaO-STO system when the moment magnetization is aligned along the $x$ or $y$ directions. 

When $\boldsymbol{M}$ is aligned along the $y$ direction, the Hamiltonian Eq. \eqref{eq:h-main} becomes invariant under the transformation $(k_x,k_y, \sigma_x,\sigma_y)\rightarrow (k_x,-k_y,-\sigma_x, \sigma_y)$. These symmetry properties imply that  $\langle \sigma_x \rangle$ is antisymmetric and $\langle \sigma_y \rangle$ symmetric about the $k_x$ axis. At the same time, $\langle v_x \rangle$ is symmetric and $\langle v_y \rangle$ antisymmetric about the same $k_x$ axis. As for the spin current $j_{i\alpha}$ where $i$ and $\alpha$ respectively denote the current flow and spin directions, the overall symmetry is given by the product of the individual symmetries of $\langle v_i \rangle$ and $\langle \sigma_\alpha \rangle$. Hence, $\langle j_{yy} \rangle$ is antisymmetric and $\langle j_{xy} \rangle$ is symmetric with respect to the $k_x$ axis. 

Let us consider the $k$-space symmetry of the first-order response. From Eq. \eqref{O1},  the first-order response is the $k$-space integral of the product of $O_\mu$ and $\delta f^{(1)}_\mu$. By considering Eq. \eqref{Ef1} and noting that in RTA, $\delta\epsilon_\mu = eE \langle v_x \rangle \tau$, $\delta f^{(1)}_\mu$ has the same symmetry as $\langle v_x \rangle$, i.e., it is symmetric about the $k_x$ axis. Therefore, the symmetry of the integrand about the $k_x$ axis in Eq. \eqref{O1} is given by the product of the symmetry of $O$ and the even symmetry of $\langle v_x \rangle$. Hence, for an $O_\mu$ that is antisymmetric about the $k_x$ axis, the integral cancels out to zero after the $k$-space integration over the entire Fermi surface. Note that the converse result does not necessarily hold, i.e., an observable $O_\mu$ that is symmetric about the $k_x$ axis will not necessarily result in a finite response (e.g. it may still integrate to zero if the $k_y$ axis is an antisymmetry axis.) For the example of $O=v_x$ shown in Fig. \ref{gFsym1o}a, the finite $M_y$ breaks the exact symmetry about the $k_y$ axis by causing an elongation of the Fermi surface along the $k_x$ axis (as can be seen from the sharper curvature of the Fermi surface at negative values of $k_x$ ), thus resulting in a finite first-order response. 

The symmetry properties of the second-order response are determined by those of its two constituent components given in Eq. \eqref{O2c}. The product $(\partial_\epsilon\delta\epsilon_\mu)(\delta f^{(1)})$ in the first term is symmetric about the $k_x$ axis because both $\partial_\epsilon\delta\epsilon_\mu$ and $\delta f^{(1)}$ have the same symmetries about the $k_x$ axis. The symmetry of the former can be seen from the following argument: Consider the expectation value of a generic observable $O$ at point $(k_x$, $k_y)$ on the Fermi surface of the $\mu$th band $O_\mu(k_x,k_y)$ where $O_\mu(k_x,k_y)=\pm O_\mu(k_x,-k_y)\ \forall k_x, k_y$. By definition, for a given energy $\epsilon$ and $k_x$, 
\begin{align}
	\partial_\epsilon O_\mu(k_x,k_y) &= \lim_{\delta\epsilon\rightarrow 0} [O_\mu\left(k_x, k_y(\epsilon+\delta\epsilon, k_x)\right) - O_\mu\left(k_x,k_y(\epsilon, k_x)\right)]/\delta\epsilon \label{eOmuSym1} \\
	&= \pm \lim_{\delta\epsilon\rightarrow 0} [O_\mu\left(k_x,-k_y(\epsilon_{\mathrm{F}}+\delta\epsilon, k_x)\right) - O_\mu\left(k_x,-k_y(\epsilon_{\mathrm{F}}, k_x)\right)]/\delta\epsilon \label{eOmuSym2} \\
	&= \pm \partial_\epsilon O_\mu(k_x,-k_y) \label{eOmuSym3}
\end{align}
where we have explicitly written $k_y$ on the Fermi surface as a function of the $\epsilon_{\mathrm{F}}$ and $k_x$. Equations \eqref{eOmuSym1}--\eqref{eOmuSym3} show that $\partial_\epsilon O_\mu$ has the same symmetry as $O_\mu$ about a symmetry axis. Since $\delta f^{(1)}_\mu$ is proportional to $\delta\epsilon_\mu$, both $(\partial_\epsilon\delta\epsilon_\mu) $ and $\delta f^{(1)}_\mu$ have the same symmetries about the $k_x$ axis. Their product is therefore symmetric. The symmetry of $(\partial_\epsilon\delta\epsilon_\mu)(\delta f^{(1)}_\mu) O_\mu$ about the $k_x$ axis is thus determined by that of $O_\mu$. We now consider the second component in Eq. \eqref{O2c}, i.e., $(\partial_\epsilon O_\mu\delta\epsilon_\mu)(\delta f^{(1)}_\mu)$. As explained, the symmetry of $\partial_\epsilon O_\mu$ about the  $k_x$ symmetry axis is the same as that of $O_\mu$ itself, while the product of the remaining $(\delta\epsilon_\mu)(\delta f^{(1)}_\mu)$ terms is symmetric (Fig. \ref{gFsym2o}a) since $\delta f^{(1)}_\mu$ is proportional to $\delta \epsilon_\mu$. Therefore, putting everything together, the symmetry of the integrand in Eq. \eqref{O2c} about the $k_x$ axis when the electric field is along the $x$ direction and in the presence of a finite $M_y$ will be the same as that of $O_\mu$. The second-order responses for $v_y$ and $\sigma_x$ are therefore antisymmetric about the $k_x$ axis and cancel out to zero. The symmetry properties of the various observables and their first- and second-order responses are summarized in Table I. 

\begin{table}
\begin{tabular}{|c|c|c|c|}
\hline
$O$ & Sym. $O$  & Sym $\delta O^{(1)}$ & Sym $\delta O^{(2)}$  \\
\hline
$v_x$ & + & + & + \\
$v_y$ & - & - & - \\
$\sigma_x$ & - & - & - \\
$\sigma_y$ & + & + & + \\
$j_{xy}$ & + & + & + \\
$j_{yy}$ & - & - & - \\ 
\hline
\end{tabular}
\label{Tab:SymMy}
\caption{Symmetries of various observables and their first- and second-order responses w.r.t. the $k_x$ symmetry axis for magnetization $\boldsymbol{M}$ and  electric field aligned along the $y$ and $x$ axes, respectively. + and - denote symmetric and anti-symmmetric, respectively. } 
\end{table}

We now consider the case where $\boldsymbol{M}$ is aligned along the $x$ direction.  The Hamiltonian Eq. \eqref{eq:h-main} is now invariant under the transformation $(k_x, k_y, \sigma_x, \sigma_y)\rightarrow (-k_x,k_y,\sigma_x, -\sigma_y)$ and the $k_y$ axis is a symmetry axis of the system (Fig. \ref{gFsym2o}b). Thus, $\langle v_y \rangle$, $\langle \sigma_x \rangle$, and $\langle j_{yx}\rangle$ are symmetric about the $k_y$ axis, while $\langle v_x \rangle$, $\langle \sigma_y \rangle$, and $\langle j_{xx}\rangle$ are antisymmetric.   In marked contrast to the $M_y$-magnetized system discussed earlier where $\langle v_x \rangle$ is symmetric about its symmetry axis along the $k_x$ axis, in the $M_x$-magnetized system, $\langle v_x \rangle$  and therefore $\delta\epsilon$, are antisymmetric about the symmetry axis along the $k_y$ axis. Therefore, the first-order responses for $O_\mu$ will have the opposite symmetries w.r.t. the $k_y$ axis compared to that of the parent $O_\mu$. However, the arguments presented in the previous paragraph for the symmetry of the second-order response of $O_\mu$ are still applicable here. Because both $(\partial_\epsilon\delta\epsilon_\mu)$ and $\delta f^{(1)}$ are \textit{anti}-symmetric about the $k_y$ symmetry axis, their product and therefore the second-order response would still be symmetric about that axis. Thus, the second-order response shares the same symmetry about the $k_y$ axis as $O_\mu$ (Fig. \ref{gFsym2o}b). The symmetry properties of various observables and their first- and second-order responses are summarized in Table II. 

\begin{table}
\begin{tabular}{|c|c|c|c|}
\hline
$O$ & Sym. $O$  & Sym. $\delta O^{(1)}$ & Sym. $\delta O^{(2)}$  \\
\hline
$v_x$ & - & + & - \\
$v_y$ & + & - & + \\
$\sigma_x$ & + & - & + \\
$\sigma_y$ & - & + & - \\
$j_{xx}$ & - & + & - \\
$j_{yx}$ & + & - & + \\ 
\hline
\end{tabular}
\label{Tab:SymMx}

\caption{Symmetries of various observables and their first- and second-order responses w.r.t the $k_y$ symmetry axis for magnetization $\mathbf{M}$ and  electric field both along the $x$ axis. + and - denote symmetric and anti-symmmetric, respectively. } 
\end{table}

Comparing the results in Tables I and II, it can be seen that for the observables listed, the first- and second-order responses have the same (opposite) symmetries with respect to the symmetry axis of $k_x$ when $\boldsymbol{M}$ is along the $y$ ($x$) direction. This implies that in the $M_x$-magnetized system, a finite first-order response for an observable is accompanied by a finite second-order response, whereas in the $M_y$-magnetized system, we can have the remarkable situation of a finite second-order response for an observable without a net first-order response.  For illustration, we present exemplary numerical calculations in the next section.

\section{Results and discussion} 

For all the numerical results that follow, we take $J_HM=10$ meV and $\epsilon_{\mathrm{F}} = 30$ meV from the band bottom of the le$^-$ band, at which Eq. \eqref{eq:h-main} is a good description of the electronic behavior of the LaO-STO system \cite{Paper1}. The SOC strengths $\alpha$, $\beta_3$, and $\eta_3$ in Eq. \eqref{eq:h-main} are treated as free parameters and varied  over a range of -10 to 10 meV\AA  for $\alpha$ and from -3 to 3 eV\AA$^{3}$ for  $\beta_3$ and $\eta_3$, based on the values in Ref. \cite{Ho_2019} for typical thicknesses and out-of-plane electric fields in LaO/STO quantum well structures. In keeping with the commonly used symbols for the various quantities, we define the charge currents $\delta J^{(1,2)}_{x,y}$, spin accumulations $\delta S^{(1,2)}_{x,y}$, and spin currents  $\delta J^{(1,2)}_{x,y;x,y}$, as $\delta O^{(1,2)}$ in Eqs. \eqref{O1} and \eqref{O2c} with $O=v_{x,y}$, $\sigma_{x,y}$, and $\frac{1}{2} \{ v_{x,y}, \sigma_{x,y} \}$, respectively. All the results that follow were obtained using the SL approach.

\begin{figure}[htp]
\centering
\includegraphics[width=0.6\textwidth]{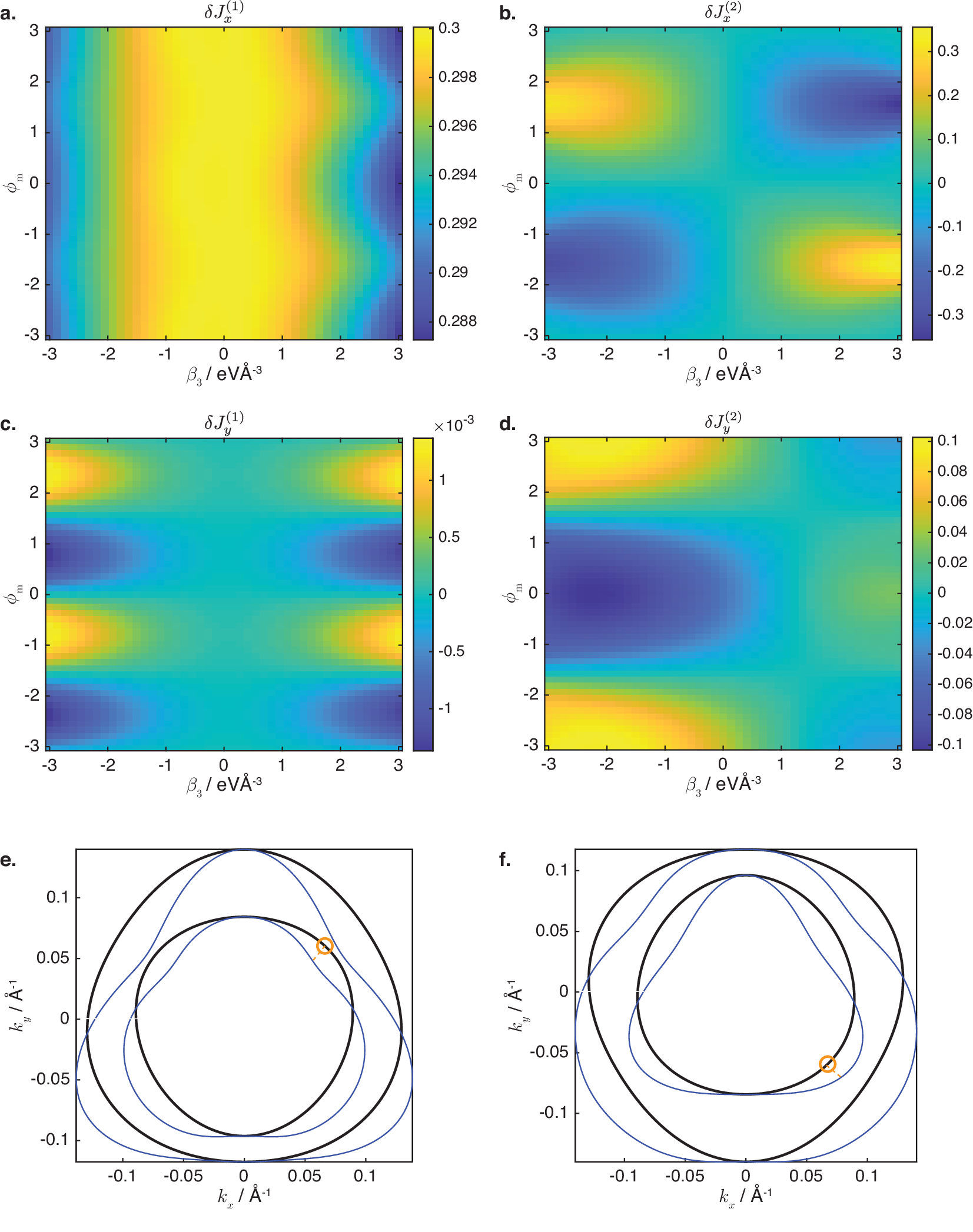}
\caption{ (a) First- ($\delta J^{(1)}_x$) and (b) second-order responses ($\delta J^{(2)}_x$) for current parallel to applied electric field, and (c) first- ($\delta J^{(1)}_y$) and (d) second- ($\delta J^{(2)}_y$) order responses in $e=\hbar=1$, eV, \AA units and unit electric field for current perpendicular to electric field as functions of the magnetization direction $\phi_{\mathrm{m}}$ and the $\beta_3$ SOC parameter for $\alpha=5$ meV\AA and $\eta_3=1$ meV\AA$^3$. (e) and (f) show the Fermi surfaces (thick black curves) and the sign and magnitude of the contribution of each state on the Fermi surfaces (thin blue lines) to $\delta J^{(2)}_y$ for magnetizations along the (e) $-x$ and (f) $+x$ directions at $\beta_3=3$ meV\AA$^3$. (The sign of the contribution of each state on the Fermi surface to  $\delta J^{(2)}_x$ is indicated by whether its corresponding point on the blue curve lies within (negative value) or outside the Fermi surface (positive value), and its relative magnitude by the separation between that point and the Fermi surface. The circles in (e) and (f) indicate a pair of states with opposite signs of the $x$ magnetization, $k_y$, and second-order contribution.  }

\label{gFig1}
\end{figure}	

Fig. \ref{gFig1} shows the first- and second-order responses for the charge current parallel ($\delta J^{(1,2)}_x$) and perpendicular ($\delta J^{(1,2)}_y$) to the applied electric field as functions of the magnetization angle ($\phi_{\mathrm{m}}$) and SOC strength $\beta_3$. In agreement with the predictions in Table II based on symmetry arguments, $\delta J^{(1)}_x$ and $\delta J^{(2)}_x$ have finite values while $\delta J^{(1)}_y$ and $\delta J^{(2)}_y$ integrate to zero when $\phi_{\mathrm{m}}=0,\pi$, which correspond to $\boldsymbol{M}$ lying along the $\pm x$ directions. Similarly, the finite values for $\delta J^{(1)}_x$ and $\delta J^{(2)}_y$ and zero net values  for $\delta J^{(1)}_y$ and $\delta J^{(2)}_x$ at $\phi_{\mathrm{m}}=\pm \pi/2$, corresponding to $\mathrm{M}$ along the $\pm y$ directions, are consistent with the the predictions in Table I. Similarly, we also evaluated the spin accumulations perpendicular to the magnetization ($\delta S^{(1/2)}_\perp$) and spin currents with spin parallel to the magnetization and flowing parallel and perpendicular to the electric field $\delta J^{(1/2)}_{(x,y)\parallel}$ as functions of $\phi_m$ and $\beta_3$. The corresponding first- and second-order responses are all in agreement with Tables I and II, as shown in the plots in Supplementary Materials Fig. S2.  

At intermediate magnetization angles between the $\pm x,y$ directions, the symmetries / antisymmetries described in Sect. \ref{sec:SymAnal} do not hold exactly. As a result, the magnitudes of the first- / second- order responses that are exactly symmetric about their respective symmetry axes when the magnetization angles are along the $\pm x$ or $\pm y$ direction would now decrease as we move away from the symmetry axes  owing to partial cancellation by the emergent antisymmetric contributions at the intermediate angles. Conversely, the first- / second-order responses that are exactly antisymmetric about their respective symmetry axes would now acquire a finite net value at the intermediate magnetization orientations due to imperfect cancellation. This gives rise to the approximately $\sin(\phi_{\mathrm{m}})$ variation of $\delta J^{(2)}_x$ in Fig. \ref{gFig1}b, which is consistent with experimental results observed in the LaO/STO system\cite{NatComm10_4510}, as well as the finite values of $\delta J^{(1)}_y$ in Fig. \ref{gFig1}c and $\delta S^{(2)}_\perp$ in Fig. S2b at intermediate values of $\phi_m$, whereas both quantities go to zero at $\phi_m=0,\pm \pi/2$ due to perfect cancellation.

An interesting point to note is that the signs of the second-order responses for the charge current in the $i=(x,y)$ direction flips when the magnetization direction is reversed about the $i$ axis, as shown in Fig. \ref{gFig1}b and d. The origin of this flip was alluded to in the discussion on Fig. \ref{gFsym2o}a, where we noted that for a finite response to be obtained for $\mathbf{M}$ along the $y$-direction, not only should the the quantity be symmetric about its $k_x$ symmetry axis, but additionally its antisymmetry about the $k_y$ axis should also be broken. This antisymmetry is broken by the applied magnetization. We show the breaking of the antisymmetry by the magnetization more explicitly in Fig. \ref{gFig1}e and f, which show the Fermi surfaces and second-order contributions to $J^{(2)}_y$ for magnetizations applied along the $-x$ and $+x$ directions, respectively. The Fermi surfaces for the two magnetization directions are reflections of each other about the $k_x$ axis. This can be shown analytically by considering the eigenvalues of Eq. \eqref{eq:h-main}, which are given by 
\begin{equation}
	\epsilon_\mu = \frac{k^2}{2m^*} \pm \sqrt{ (J_HM_x + k_y(\alpha-k_y^2\beta_3+k_x^2(\beta_3-\eta_3)))^2 + (J_H M_y - k_x(\alpha + k_x^2\beta_3 - k_y^2(\beta_3-\eta_3))^2} \label{egVl},
\end{equation}
and is invariant under $(M_x,k_y)\rightarrow (-M_x,-k_y)$ when $M_y=0$. For simplicity, we denote the eigenvalue as $\epsilon_\mu^\pm$ when  $M_y=0$ and $M_x=\pm |M_x|$. Thus, we have $\epsilon_\mu^-(k_x,k_y) = \epsilon_\mu^+(k_x,-k_y)$. By definition, $\langle v_y^\pm (k_x, k_y') \rangle = \partial_{k_y'}\epsilon^\pm_\mu(k_x, k_y')$. Setting $k_y' = \pm k_y$ yields $\langle v_y^\pm (k_x,k_y) \rangle = \pm \langle v_y^+ (k_x,\pm k_y) \rangle$. (More intuitively, this can be seen by noting that for states that are particle-like, the Fermi surfaces become larger with increasing energy so the states with large positive (negative) values of $k_y$ extend upwards with increasing energy and have positive (negative) expectation values for $v_y$ regardless of whether the magnetization is along the $+x$ or $-x$ directions.)  Therefore, the contribution to the second-order response, $\delta J_y^{(2)}$, of a state at $(k_x,k_y)$ on the Fermi surface in the $+x$ magnetized system  has the same magnitude but opposite sign to that of the corresponding state at $(k_x,-k_y)$ in the $-x$ magnetized system (compare Fig. \ref{gFig1}e and f). The net resultant value of $\delta J^{(2)}_y$ therefore has the same magnitude but opposite signs when the magnetization direction switches from $+x$ to $-x$ after integrating over the entire Fermi surface and summing up over both bands. (Although the argument provided pertains to only the $\pm x$ magnetization directions,  it can be generalized to magnetization directions along any arbitrary axis.)

\begin{figure}[htp]
\centering
\includegraphics[width=0.9\textwidth]{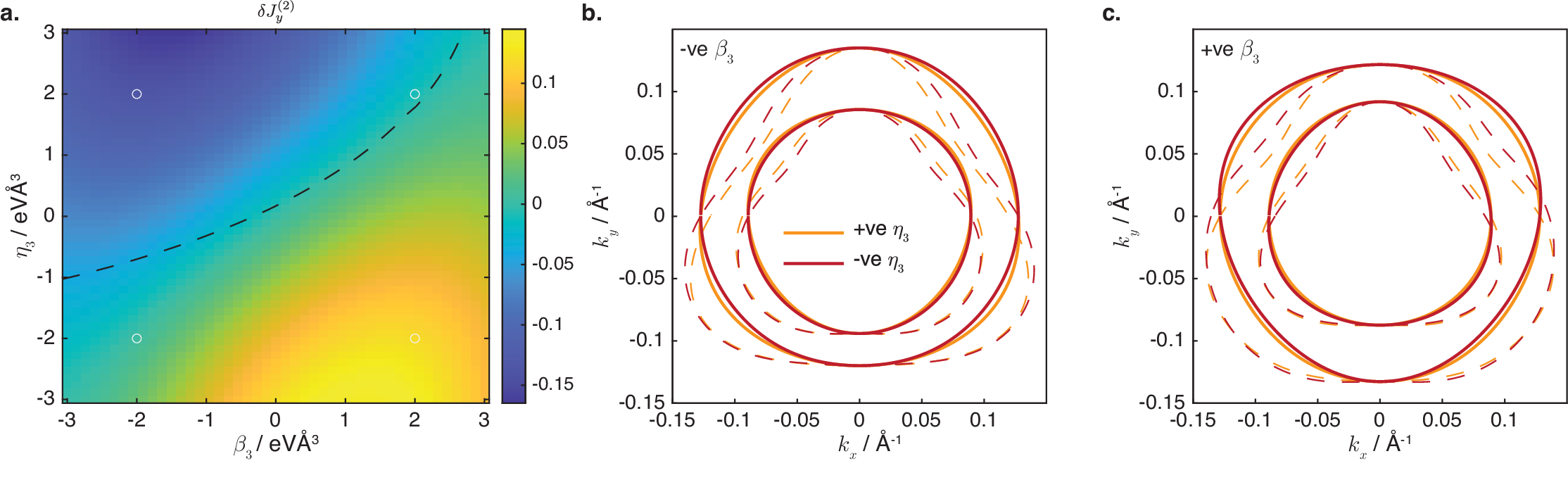}
\caption{ (a) Variation of $\delta J^{(2)}_y$ with $\eta_3$ and $\beta_3$ at $\alpha=5$ meV\AA and $\phi_{\mathrm{m}}=0$. The dotted line traces the loci of $\eta_3$ an $\beta_3$ at which $\delta J^{(2)}_y = 0$. (b) and (c) show the Fermi surfaces and the sign and magnitudes of the relative contributions for each state on the Fermi surfaces for $\eta_3=\pm 2$ meV\AA and (b) $\beta_3=-2$ meV\AA$^3$ and (c) $\beta_3= 2$ meV\AA$^3$.  The sign of the contribution of each state on the Fermi surface to  $\delta J^{(2)}_x$ is indicated by whether its corresponding point on the dotted line of the same color lies within (negative value) or outside the Fermi surface (positive value), and its relative magnitude by the separation between that point and the Fermi surface. The points $\eta_3=\pm 2$ meV\AA$^3$ and $\beta_3=\pm 2$ meV\AA$^3$ depicted in panels (b) and (c) are marked by the white circles in (a). }

\label{gFig3}
\end{figure}


We next consider the effect of the cubic SOC terms on the second-order Hall charge current response when the magnetization is parallel to the electric field. (The first-order charge current is zero, as shown in Table I.) Fig. \ref{gFig3}a shows that the sign and magnitude of $\delta J^{(2)}_y$ can be under varying $\eta_3$ and / or $\beta_3$. As discussed earlier, the net value of $\delta J^{(2)}_y$ is given by the sum of contributions from the states integrated over the Fermi surfaces of the two bands, as shown in Fig. \ref{gFig3}b and c. Because the contributions from each state does not all have the same sign, the sign of the net $\delta J^{(2)}_y$ after summing over all states depends on the balance between the positive and negative contributions on the Fermi surfaces. This in turn is determined by (i) the magnitude of the contribution of each state and (ii) the portions on the Fermi surfaces occupied by states with positive and negative contributions. For example, at $\beta_3=-2$ meV\AA$^3$, $\delta J^{(2)}_y$ is negative at $\eta_3=2$ meV\AA$^3$ and positive at $\eta_3=-2$ meV\AA$^3$ as shown in Fig. \ref{gFig3}a. The Fermi surfaces and contributions of each state to $\delta J^{(2)}_y$ corresponding to these values of $\beta_3$ and $\eta_3$ are shown in Fig. \ref{gFig3}b. We focus on the outer Fermi surface, which has a larger $k$-space length corresponding to more states and therefore plays a larger role in determining the net value of $\delta J^{(2)}_y$. The negative value of $\delta J^{(2)}_y$ at $\eta_3=2$ meV\AA$^3$ can then be explained by the larger magnitude of the negative contributions from states at the top half of the Fermi surface  (note that the dark yellow dotted line for $\eta_3=2$ meV\AA$^3$ cuts a deeper recess into the Fermi surface), as well as the larger portion of the Fermi surface occupied by states with negative contributions (the contributions switch to positive at a larger value of $k_y$ compared to the corresponding case of $\eta_3=-2$ meV\AA$^3$). Fig. \ref{gFig3}c shows a complementary example for $\beta_3=2$ meV\AA$^3$ where the magnitude of the negative $\delta J^{(2)}$ is smaller than that at $\beta_3=-2$ meV\AA$^3$ keeping the other cubic SOC parameter fixed at $\eta_3=2$ meV\AA$^3$. The decrease in the magnitude of $\delta J^{(2)}_y$ in the former can be attributed to the fact the portions of the outer Fermi surface with negative contributions now constitute a smaller proportion of the total length of the Fermi surface.

Lastly, the plot in Fig. \ref{gFig3}a indicates that $\eta_3$ has a larger influence on the sign of $\delta J^{(2)}_y$ than $\beta_3$. (This can be seen from the fact that the dotted line denoting the locus of $\delta J^{(2)}_y$ lies closer to the horizontal line, which indicates that the sign of $\delta J^{(2)}_y$ is independent of $\beta_3$, than the vertical line, which indicates that the sign of $\delta J^{(2)}_y$ is independent of $\eta_3$. )  The larger influence of $\eta_3$ can also be seen by rewriting the energy eigenvalues in Eq. \eqref{egVl} in polar coordinates
\begin{align}
	\epsilon_\mu =& \frac{k^2}{2m^*} \pm \left( (- k\alpha\sin(\phi) + k^3(\beta_3+\eta_3)\cos^2(\phi)\sin(\phi)  - k^3\sin^3(\phi)) \right. \nonumber \\
	&+ \left. (J_HM_x + k\alpha\cos(\phi) + k^3(\beta_3-\eta_3)\cos(\phi)\sin^2(\phi) - k^3\beta^3\cos(\phi)^3) \right)^{\frac{1}{2}},
\label{egVlPol} 
\end{align}  
from which it can be seen that the terms containing $\eta_3$ have the largest magnitudes at $\phi=\pm \pi/4, \pm 3\pi/4$. These values of $\phi$ coincide with those at which the contributions to $\delta J^{(2)}_y$ have the largest magnitudes. Note that the contributions to  $\delta J^{(2)}_y$ go to zero at $\phi=\pm \pi/2$ since $\langle v_x \rangle = 0$, which leads to $\delta\epsilon_\mu=0$ by Eq. \eqref{O2c}, and also at $\phi=0,\pi$ where $\langle v_y \rangle=0$). Therefore, changing the value of $\eta_3$ would cause a relatively large change in the contribution to $\delta J^{(2)}_y$. At the same time, changing the sign of $\eta_3$ also has the significant effect of changing the shape of the Fermi surfaces and therefore, the relative lengths of the portions with positive and negative contributions to $\delta J^{(2)}_y$  (see Fig. \ref{gFig3}b and c.) This is compatible with experimentally findings that reveal the sensitivity of the second-order response in WTe$_2$ to the shape of the Fermi surface \cite{NatComm10_1290}.
  
\section{Conclusions}
In this study, we investigated the second-order spin accumulation and spin and charge current responses to an applied electric field in a LaO/STO system in the presence of cubic SOC terms and magnetic dopants. We first extended the approach of Schliemann and Loss for solving the Boltzmann transport equation to higher orders in the applied electric field. We then explained the physical origin of the second-order response of an observable quantity as the combined effects of the change in the expectation value due to the Fermi surface shift and the energy dependence of the shift. Subsequently, we performed a symmetry analysis on the LaO/STO system when the magnetization of the dopants lies parallel or perpendicular to the applied electric field. We showed how the interplay between the magnetization direction and the shift in the Fermi surface shift leads to antisymmetrical distributions of some of the observable quantities, which results in the cancellation of their corresponding first- or second-order responses in some cases. Conversely, in other cases broken antisymmetry results in finite responses . These theoretical predictions were subsequently confirmed by numerical calculations. Finally, we further explored the magnetization direction dependence of the first- and second-order responses and the interplay between the two cubic RSOC strength. We showed that the sign of the second-order responses can be switched by varying either the magnetization direction or relative magnitudes of the cubic RSOC terms, a finding that may be explained by the $k$-space symmetry of the Fermi surfaces of the system. These findings extend our understanding of how the spin and charge responses are affected by the interplay between multiple SOC effects in a LaO/STO system, and how the relative magnitudes of the first- and second-order responses of various quantities can be engineered for practical applications by exploiting the symmetries of the system. 

\section*{Acknowledgments}
This work is supported by the Ministry of Education (MOE) Tier-II Grant No. MOE-T2EP50121-0014 (NUS Grant No. A-8000086-01-00), and MOE Tier-I FRC grant (NUS Grant No. A-8000195-01-00).

\bibliographystyle{unsrt}
\bibliography{Bibtex-LAO-STO-v2}

\end{document}


\author{Zhuo Bin Siu}
\email{elesiuz@nus.edu.sg}
\affiliation{Department of Electrical and Computer Engineering, National University of Singapore, Singapore 117583, Republic of Singapore}
\author{Anirban Kundu}
\email{anirbank5@gmail.com}
\affiliation{Department of Electrical and Computer Engineering, National University of Singapore, Singapore 117583, Republic of Singapore}
\affiliation{Department of Physics, Ariel University, Ariel 40700, Israel}
\author{Mansoor B.A. Jalil}
\email{elembaj@nus.edu.sg}
\affiliation{Department of Electrical and Computer Engineering, National University of Singapore, Singapore 117583, Republic of Singapore}
\title{Second-order charge and spin transport in LaO/STO system in the presence of multiple Rashba spin orbit couplings}
\maketitle
\renewcommand{\figurename}{Supplementary Figure}

\begin{figure}[htp]
\centering
\includegraphics[width=0.85\textwidth]{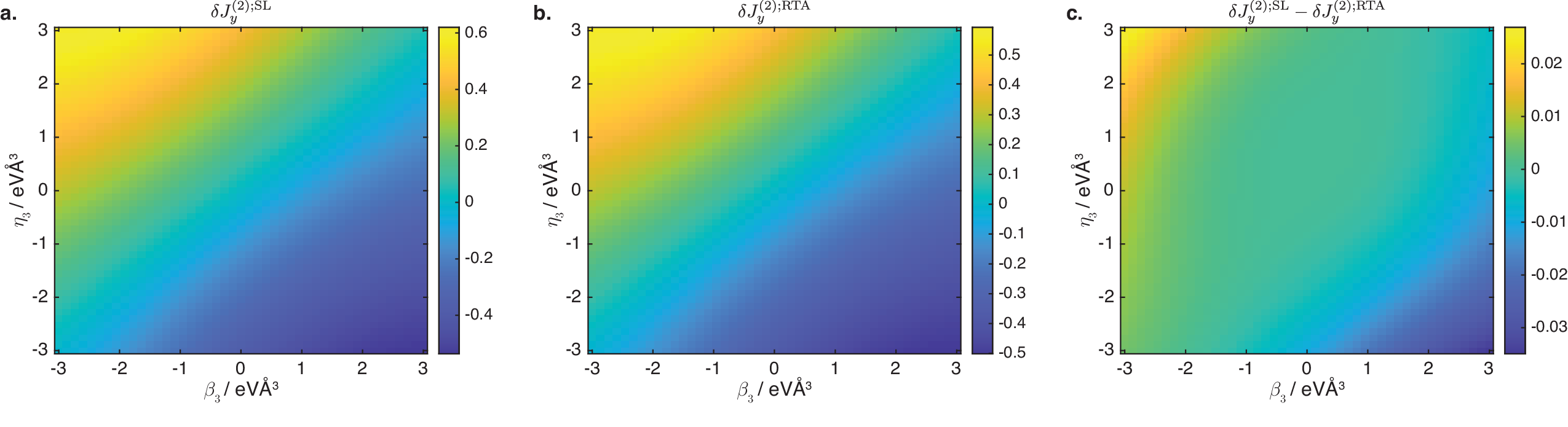}
\caption{ (a), (b) The second-order charge current perpendicular to the electric field $\delta J^{(2)}_y$ at $\alpha=0.05$ meV\AA and $\phi_{\mathrm{m}}=0$ as a function of $\beta_3$ and $\eta_3$ obtained using the (a) SLA and (b) RTA, and (c) their differences.   }
\label{SF3}
\end{figure}	

Fig. \ref{SF3} shows a comparison between the second-order charge current flowing perpendicular to the applied electric field calculated using the Schliemann-Loss and relaxation-time approximation approaches, in which the relaxation time was taken as the mean value of $\tau^\parallel_\mu$ over the Fermi surfaces of both bands. The RTA produces qualitatively similar results to the Schliemann-Loss approach but underestimates the magnitudes of the second-order response.

\begin{figure}[htp]
\centering
\includegraphics[width=0.6\textwidth]{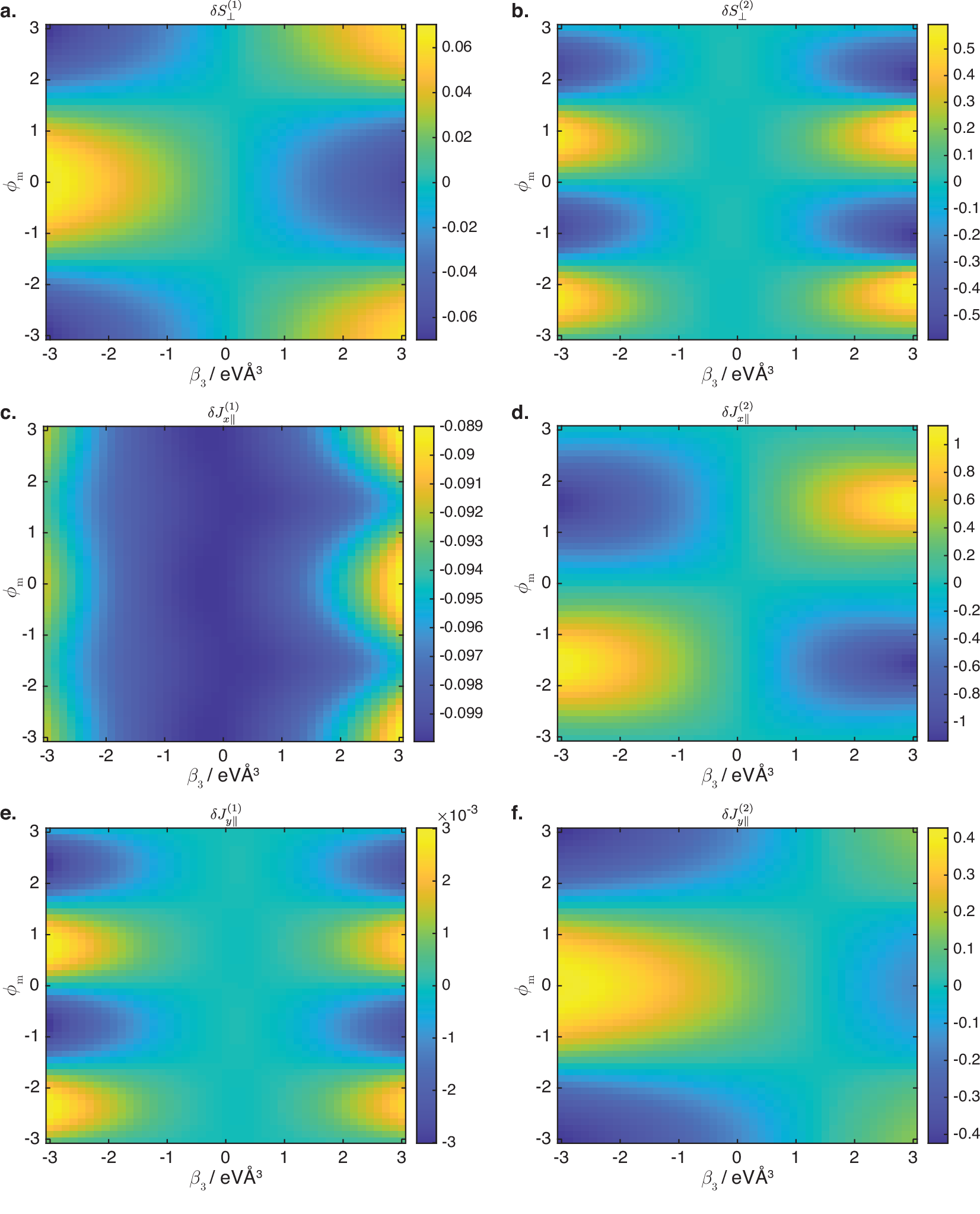}
\caption{ The (a, c, e) first-order responses for the (a) spin accumulation perpendicular to the magnetization direction $\delta S^{(1)}_\perp$, (b) spin current with spin polarization parallel to the magnetization direction flowing parallel to the applied electric field $\delta J^{(1)}_{x\parallel}$, (c) and spin current parallel to the magnetization direction flowing perpendicular to the applied electric field $\delta J^{(1)}_{y\parallel}$, and (b, d, f) the corresponding second-order responses  as functions of the magnetization direction $\phi_{\mathrm{m}}$ and the $\beta_3$ SOC parameter for $\alpha=5\ \mathrm{meV}$\AA and $\eta_3=1\ \mathrm{meV}$\AA. }
\label{SF1}
\end{figure}	

Fig. \ref{SF1} shows the first- and second-order responses for the spin accumulations perpendicular to the magnetization direction and the spin currents with spin polarizations parallel to the magnetization directions flowing parallel and perpendicular to the applied electric field. The results at $\phi_{\mathrm{m}}=0,\pi$ and at $\phi_{\mathrm{m}}=\pm pi/2$ are consistent with Tables I and II in the main manuscript, respectively. Interestingly, a finite second-order spin accumulation perpendicular to the magnetization directions at intermediate magnetization angles between the $\pm x$ and $\pm y$ directions emerges in Fig. \ref{SF1}b.

\begin{figure}[htp]
\centering
\includegraphics[width=0.6\textwidth]{SuppFig2.eps}
\caption{ The second-order responses for the charge current perpendicular to the applied electric field $\delta J^{(2)}_y$ as functions of (a), (b) $\alpha$ and $\eta_3$ at (a) $\beta_3=-2$ eV\AA$^3$ and (b)  $\beta_3=2$ eV\AA$^3$, and as functions of (c), (d) $\alpha$ and $\beta_3$ at (c) $\eta_3=-1$ eV\AA$^3$ and (d) $\eta_3=1$ eV\AA$^3$. The black dotted line in each plot denotes the locus of the points where $\delta J^{(2)}_y=0$.  }
\label{SF2}
\end{figure}	

Fig. \ref{SF2} shows the variation of $\delta J^{(2)}_y$ with $\alpha$, $\beta_3$, and $\eta_3$. For the parameter ranges in this studyt, $\alpha_3$ does not have a substantial effect on the sign of $\delta J^{(2)}_y$, as indicated by the fact that the dotted black lines denoting the points where $\delta J^{(2)}_y=0$ are almost vertical. Fig. \ref{SF2}a and b show that the sign of $\delta J^{(2)}_y$ can be switched by varying $\eta_3$, and Fig. \ref{SF2}c and d show that the sign of $\delta J^{(2)}_y$ can be switched by varying $\beta_3$.

\bibliographystyle{unsrt}
\bibliography{Bibtex-LAO-STO-v2}